\documentclass[a4paper]{article}

\usepackage{INTERSPEECH2020}
\usepackage{subfigure}
 \usepackage{cite} 
\title{Speech Separation Based on Multi-Stage Elaborated Dual-Path Deep BiLSTM with Auxiliary Identity Loss
}
\name{Ziqiang Shi$^1$, Rujie Liu$^1$, Jiqing Han$^2$}
\address{
  $^1$Fujitsu Research and Development Center\\
  $^2$Harbin Institute of Technology}
\email{shiziqiang@cn.fujitsu.com}

\begin{document}

\maketitle
\begin{abstract}
Deep neural network with dual-path bi-directional long short-term memory (BiLSTM) block has been proved to be very effective in sequence modeling, especially in speech separation. This work investigates how to extend dual-path BiLSTM to result in a new state-of-the-art approach, called TasTas, for multi-talker monaural speech separation (a.k.a cocktail party problem). TasTas introduces two simple but effective improvements,  one is an iterative multi-stage refinement scheme, and the other is to correct the speech with imperfect separation through a loss of speaker identity consistency between the separated speech and original speech, to boost the performance of dual-path BiLSTM based networks. TasTas takes the mixed utterance of two speakers and maps it to two separated utterances,  where each utterance contains only one speaker's voice. Our experiments on the notable benchmark WSJ0-2mix data corpus result in 20.55dB SDR improvement, 20.35dB SI-SDR improvement, 3.69 of PESQ, and 94.86\% of ESTOI, which shows that our proposed networks can lead to big performance improvement on the speaker separation task. We have open sourced our re-implementation of the DPRNN-TasNet here\footnotemark[1], and our TasTas is realized based on this implementation of DPRNN-TasNet, it is believed that the results in this paper can be  reproduced with ease.
\end{abstract}

\footnotetext[1]{https://github.com/ShiZiqiang/dual-path-RNNs-DPRNNs-based-speech-separation}

\section{Introduction}

Multi-talker monaural speech separation has a vast range of applications. For example, a home environment or a conference environment in which many people talk, the human auditory system can easily track and follow a target speaker's voice from the multi-talker's mixed voice. In this case, a clean speech signal of the target speaker needs to be separated from the mixed speech to complete the subsequent recognition work. Thus it is a problem that must be solved in order to achieve satisfactory performance in speech or speaker recognition tasks. The difficulty in this problem is that since we don't have any prior information of the user, a practical system must be speaker-independent. 

Recently, a large number of techniques based on deep learning are proposed for this task. These methods can be briefly grouped into two categories: time-frequency (TF) domain methods (non-end-to-end) and time-domain methods (end-to-end). The first category is to use short-time Fourier transform (STFT) to decompose the time-domain mixture into the time-frequency domain to display and to separate therein. Usually, deep neural networks (DNN) is introduced for estimating the ideal binary or ratio masks (IBM or IRM), or phase-sensitive masks (PSM), and the source separation is transformed into a magnitude domain TF unit-level classification or regression problem, and mixed phases are usually retained for resynthesis. Notable work includes deep clustering (DPCL)~\cite{hershey2016deep,isik2016single}, permutation invariant training (PIT)~\cite{yu2017permutation}, and combinations of DPCL and PIT, such as Deep CASA~\cite{liu2019divide} and Wang et al.~\cite{wang2019deep}.The second category is end-to-end speech separation in time-domain~\cite{luo2017tasnet,luo2018tasnet,venkataramani2017adaptive,zhang2020furcanext,luo2019dual}, which is a natural way to overcome the obstacles of the upper bound source-to-distortion ratio improvement (SDRi) in STFT mask estimation based methods and real-time processing requirements in actual use.

\begin{figure*}[th]
\centering
\subfigure[The operation of `Segmentation'.]{
\begin{minipage}[t]{0.33\linewidth}
\centering
\includegraphics[width=1.5in]{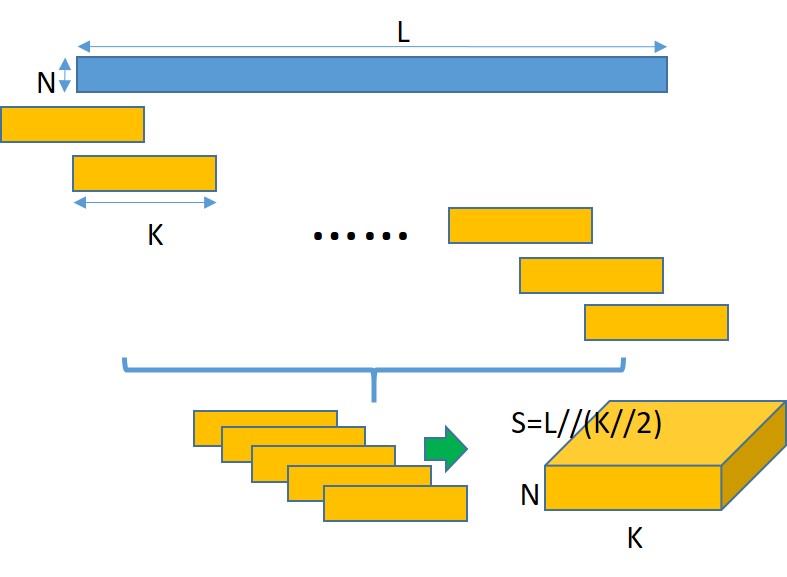}
\label{segmentation}
\end{minipage}%
}%
\subfigure[The structure of dual-path BiLSTM.]{
\begin{minipage}[t]{0.33\linewidth}
\centering
\includegraphics[width=2.3in]{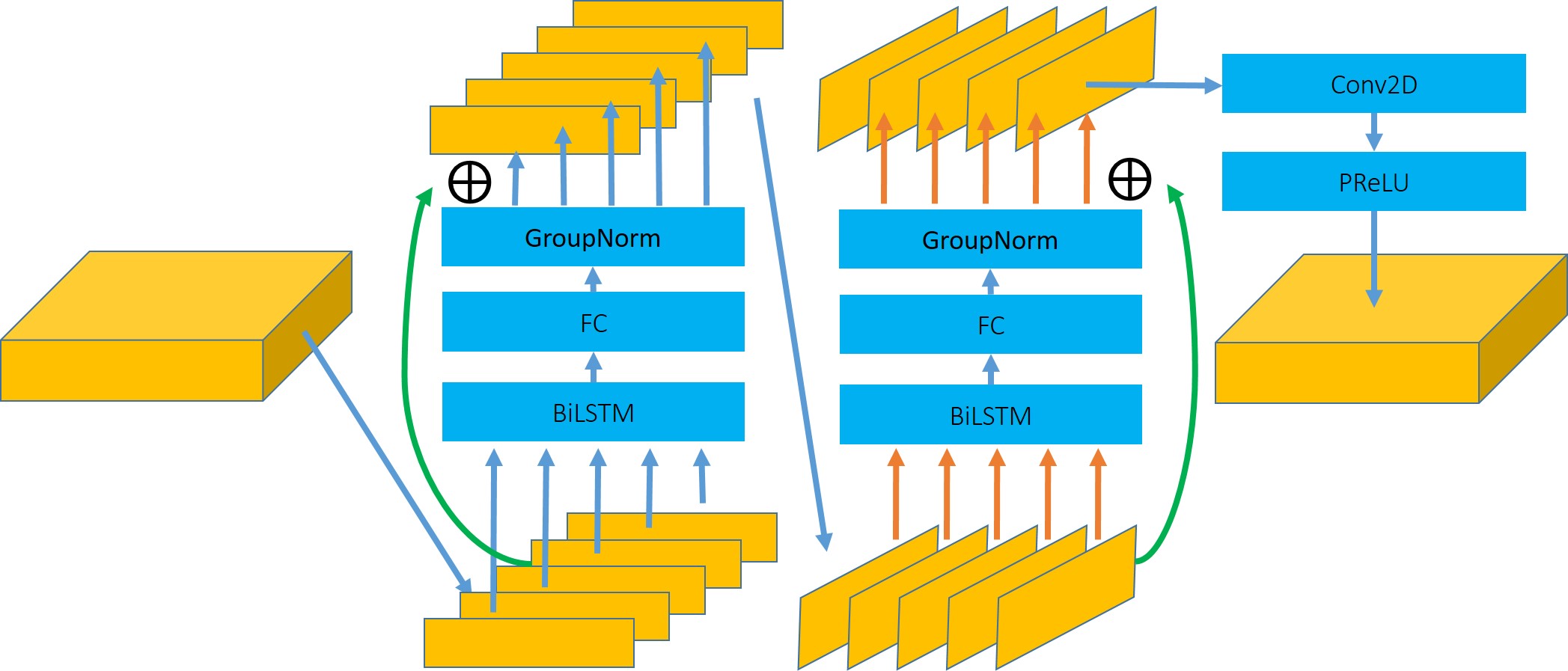}
\label{dprnn}
\end{minipage}%
}%
\subfigure[The operation of `Merge'.]{
\begin{minipage}[t]{0.33\linewidth}
\centering
\includegraphics[width=1.5in]{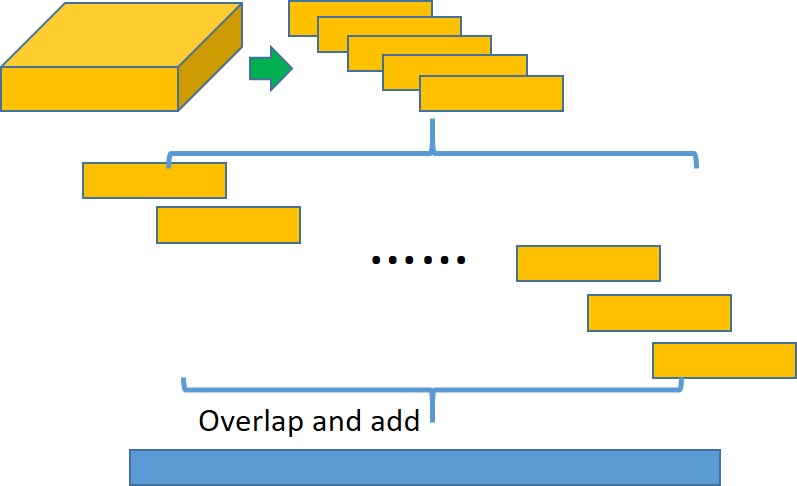}
\label{Merge}
\end{minipage}
}%
\centering
\caption{Key components in the pipeline of DPRNN-TasNet}
\end{figure*}

This paper is based on the end-to-end method~\cite{luo2017tasnet,luo2018tasnet,zhang2020furcanext,luo2019dual}, which has achieved better results than DPCL based or PIT based approaches. Since most DPCL and PIT based methods use STFT as front-end. Specifically, the mixed speech signal is first transformed from one-dimensional signal in time domain to two-dimensional spectrum signal in TF domain, and then the mixed spectrum is separated to result in spectrums corresponding to different source speeches by a clustering or mask estimation method, and finally, the cleaned source speech signal can be restored by an inverse STFT on each spectrum. This framework has several limitations. Firstly, it is unclear whether the STFT  is optimal transformation of the signal for speech separation~\cite{shi2019cqt}. Secondly, most STFT based methods often assumed that the phase of the separated signal to be equal to the mixture phase, which is generally incorrect and imposes an obvious upper bound on separation performance by using the ideal masks. As an approach to overcome the above problems, several speech separation models were recently proposed that operate directly on time-domain speech signals~\cite{luo2017tasnet,luo2018tasnet,zhang2020furcanext,luo2019dual}. Based on these first results, and inspired by~\cite{isik2016single,kavalerov2019universal,luo2019dual}, we propose TasTas, a multi-stage iterative elaborated dual-path BiLSTM based end-to-end speech separation network with an auxiliary speaker identity loss, in which the signal estimates from an initial mask-based separation network serves as input, along with the original mixture, to a next identical separation network.


\section{Speech Separation with Dual-Path BiLSTM Blocks}
\label{sec:dplstm}

In this section, we review the original dual-path BiLSTM based separation architecture~\cite{luo2019dual}. Luo et al.~\cite{luo2018tasnet,luo2019dual} introduce adaptive front-end methods to achieves high speech separation performance on WSJ0-2mix dataset~\cite{hershey2016deep,isik2016single}. Such methods contain three processing stages, here the state-of-the-art architecture~\cite{luo2019dual} is used as an illustration. The architecture consists of an encoder (Conv1d is followed by a PReLU), a separator (consisted of the order by a LayerNorm, a 1$\times$1conv, 6 dual-path BiLSTM layers, 1$\times$1conv, and a softmax operation) and a decoder (an FC layer).  First, the encoder module is used to convert short segments of the mixed waveform into their corresponding representations. Then, the representation is used to estimate the multiplication function (mask) of each source and each encoder output for each time step. The source waveform is then reconstructed by transforming the masked encoder features using a linear decoder module. This framework is called DPRNN-TasNet in~\cite{luo2019dual}.

The key factors for the best performance of DPRNN-TasNet are the local and global data chunk formulation in the dual-path BiLSTM module~\cite{luo2019dual}. Luo et al.~\cite{luo2019dual} first splits the output of the encoder into chunks with or without overlaps and concatenates them to form a 3-D tensor, as shown in Figure~\ref{segmentation}. The dual-path BiLSTM modules will map these 3-D tensors to 3-D tensor masks, as shown in Figure~\ref{dprnn}. The output 3-D tensor masks and the original 3-D tensor are converted back to a sequential output by a `Merge' operation as shown in Figure~\ref{Merge}.


Although DPRNN-TasNet has achieved a good SDR improvement~\cite{fevotte2005bss,vincent2006performance} in some public data sets, there is a clear disadvantage in this structure, that is, all consecutive frames in the input of inter-BiLSTM are far apart in the original utterance. There are few sequence information and relationship between the adjacent frames in the input of inter-BiLSTM. If the context information or mechanism can be added to the neighboring frames or to the structure of the inter-BiLSTM respectively, it is believed the performance will be boosted. At the same time, in the training of DPRNN-TasNet, the performance variance of different episodes is large, so some ensemble methods are tried to strengthen DPRNN-TasNet. Also, since the separated outputs and mixed input of the speech separation network must meet a consistent condition, that is, the sum of the separated outputs must be consistent with the mixed input. Therefore, this consistent condition can also be used to refine the separated outputs of the network. That is, the output of the DPRNN-TasNet can be refined again by combining the original mixed utterance to feed into the DPRNN-TasNet to result in better SDRi. These are the motivations for the two simple but effect improvements proposed in this section. The end-to-end speech separation network based on these two improved methods is called TasTas in this paper.

\begin{figure*}[th]
\centering
\hspace{-5mm}
\includegraphics[width=1.0\linewidth]{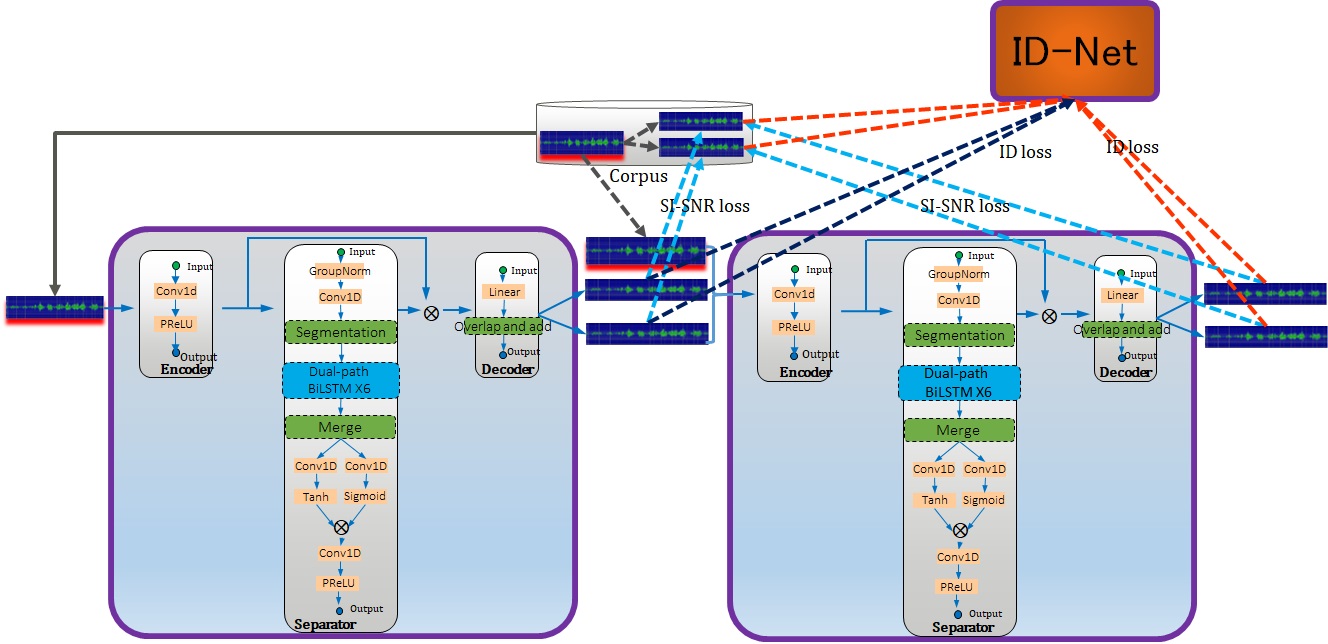}
\hspace{-5mm}
\caption{
The structure of iterative multi-stage elaborated dual-path BiLSTM with explicit speaker-aware loss for speech separation, 
which is also called TasTas.
}
\label{iterative_la_furca}
\end{figure*}

\section{Speech Separation with TasTas}
\label{sec:lafurca}

\begin{figure}[ht]
\centering
\subfigure[The structure and training method of ID-Net.]{
\begin{minipage}[t]{1\linewidth}
\centering
\includegraphics[width=2.3in]{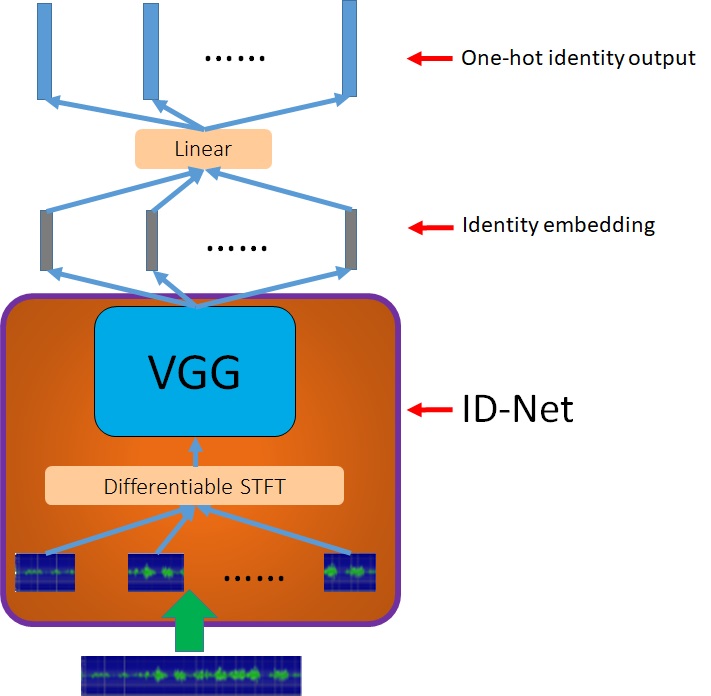}
\label{id_net_training}
\end{minipage}%
}%

\subfigure[The usage of ID-Net]{
\begin{minipage}[t]{1\linewidth}
\centering
\includegraphics[width=2in]{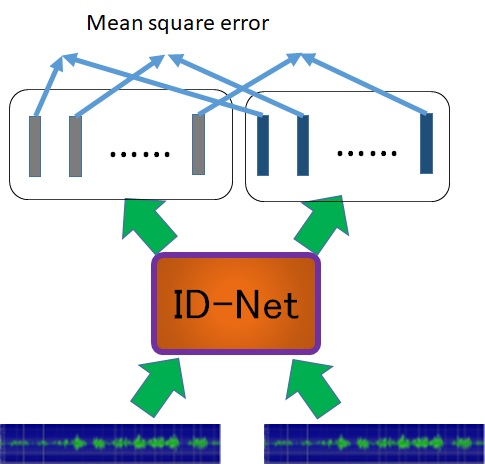}
\label{id_net_inference}
\end{minipage}
}%
\centering
\caption{The structure and usage of ID-Net.}
\label{id_net}
\end{figure}

The first improvment is to introduce an iterative multi-stage refinement scheme.
Inspired by~\cite{isik2016single,kavalerov2019universal}, as shown in Fig.~\ref{iterative_la_furca} we propose to use a multi-stage iterative network to do monaural speech separation. In each stage, there is a complete separate pipeline mentioned earlier, such as any DPRNN-TasNet. The output of each stage pipeline is two separate utterances, and these two utterances will be sent to the next stage sub-network along with the original mixed utterance to continue through the exact same pipeline, such as DPRNN-TasNet, except that the input dimension is tripled. 

Here we take an example of a two-stage TasTas in Fig.~\ref{iterative_la_furca}. During training, the original input is mixed speech, and the output is the separated utterances, which are hoped to be as close to the original clean utterances as possible. This two-stage TasTas consists of two almost identical subnetworks connected in sequence. The outputs of the first subnetwork are indeed two separated utterances, which will be compared with the clean utterances in the corpus, and the loss is calculated. Then these separated utterances will be refined by the second subnetwork. The two separated utterances are concatenated with the original mixed voice in the corpus and sent to the second subnetwork. In other words, the number of representations of the encoder output in the second subnetwork has tripled, and the dimensions of the input of subsequent separator and decoder have also tripled. Finally, the outputs of the second subnetwork are compared with the clean utterances and the loss is calculated. In our training process, both losses of the two subnetworks will be calculated and averaged as the final loss.

In our implementation and experiments, we tried different numbers of stages, including 2 stages and 3 stages. In other words, as shown in Fig.~\ref{iterative_la_furca}, 2 or 3 DPRNN-TasNets with intra-parallel BiLSTM and inter-parallel BiLSTMs are connected in sequence to form an iterative refinement network. The insight we got was that 3 or more stages did not improve the performance anymore. That is, using only two stages is enough. When using three stages, the separation performances in SDR of the first stage and the second stage are basically the same, in other words, one of the first two stages is not working. 

The second improvment is to introduce an identity network (ID-Net) to make further polish of separated utterances from the separation pipeline. This idea is motivated by~\cite{nachmani2020voice}. It is expected that the separation utterances not only maintain a high SI-SDR, but also ensure the consistency between the speaker identities from the separated speech and the original speech. As shown in Fig.~\ref{id_net_training}, the ID-Net here is obtained by connecting a differentiable STFT module\footnotemark[2] and a VGG11 network~\cite{simonyan2014very}. The ID-Net itself needs to be trained separately through the training data of WSJ0-2mix, which indeed has speaker identity information. Each utterance of training data is divided into 0.5 seconds, sent to ID-Net, and output one-hot speaker identity, as shown in Fig.~\ref{id_net_training}. After training the ID-Net, it is fixed and only the output of the penultimate layer of ID-Net is used as the speaker's identity feature vector. Make the separated utterances correspond to the original utterances one by one (after the permutation has been optimized by the separation pipeline with PIT training) ,  and feed them into the ID-Net to extract the speaker identity feature vector. The speaker identity related loss (hereinafter referred to as ID-loss) are caculated as the mean square distances between the speaker identity feature vectors. At the same time, the SI-SDR related loss  is added together for training.

\footnotetext[2]{https://github.com/pseeth/torch-stft/tree/master/torch\_stft}

\subsection{Utterance-Level Scale-Invariant SDR Objective Loss}
\label{sec:loss}

In this work, we directly use the scale-invariant signal-to-distortion ratio (SI-SDR)~\cite{fevotte2005bss,vincent2006performance, roux2018sdr}.
SI-SDR captures the overall separation quality of the algorithm. There is a subtle problem here. We first concatenate the outputs of  TasTas  into a complete utterance and then compare with the input full utterance to calculate the SI-SDR in the utterance level instead of calculating the SI-SDR for one frame at a time. These two methods are very different in ways and performance. If we denote the output of the network by $s$, which should ideally be equal to the target source $x$, then SI-SDR can be given as~\cite{fevotte2005bss,vincent2006performance,roux2018sdr}
\begin{equation*}
 \tilde{x}=\frac{\langle x , s \rangle}{\langle x , x \rangle} x, \quad e=\tilde{x}-s,\quad \text{SDR} = 10*\text{log}_{10}\frac{\langle \tilde{x} , \tilde{x} \rangle}{\langle e , e \rangle}.
\end{equation*}
Then our target is to maximize SI-SDR or minimize the negative SI-SDR as loss function respect to the $s$.

To solve the tracing and permutation problem, the PIT training criteria~\cite{yu2017permutation} is employed in this work. We calculate the SI-SDRs for all the permutations, pick the maximum one, and take the negative as the loss. It is called the SI-SDR loss in this work. The SI-SDR losses of the separated speech outputs at all stages with ground truth will be calculated, and then be averaged as the final loss.

\subsection{Training}
\label{sec:adam}

During training Adam~\cite{kingma2014adam} serves as the optimizer to minimize the SDR loss with an initial learning rate of 0.001 and scale down by 0.98 every two epochs. when the training loss increased on the development set, then restart training from the current best checkpoint with the halved initial learning rate. In other words, the learning rates of restart training are 0.001, 0.0005, 0.00025, etc. respectively. Due to the limitation of GPU memory, the batch size is set to 1, 2, or 3 according to the size of GPU

There are three phases in training TasTas. First, train the ID-Net with the paired original utterances and speaker identity information, and after the training is sufficient, the ID-Net will be fixed. Then we train the separation pipeline of TasTas without considering ID-loss, that means with only SI-SDR loss. Finally, we fine-tune the separation pipeline through both ID-loss and SI-SDR loss to complete the training.

\section{Experiments}
\label{sec:experiments}

\subsection{Dataset and Neural Network}
\label{ssec:dataset}

We evaluated our system on the two-speaker speech separation problem using the WSJ0-2mix dataset~\cite{hershey2016deep,isik2016single},  which is a benchmark dataset for two-speaker mono speech separation in recent years, thus most of those methods are compared on this dataset. WSJ0-2mix contains 30 hours of training and 10 hours of validation data. The mixtures are generated by randomly selecting 49 male and 51 female speakers and utterances in the Wall Street Journal (WSJ0) training set si\_tr\_s, and mixing them at various signal-to-noise ratios (SNR) uniformly between 0 dB and 5 dB (the SNRs for different pairs of mixed utterances are fixed by the scripts provided by~\cite{hershey2016deep,isik2016single} for fair comparisons). 5 hours of evaluation set is generated in the same way, using utterances from 16 unseen speakers from si\_dt\_05 and si\_et\_05 in the WSJ0 dataset. 

We evaluate the systems with the SDRi~\cite{fevotte2005bss,vincent2006performance}, perceptual evaluation of speech quality (PESQ)~\cite{rix2001perceptual} and extend short-time objective intelligibility (ESTOI)~\cite{jensen2016an} metrics used in~\cite{isik2016single,luo2018speaker, chen2017deep,liu2019divide,wang2019deep}. The original SDR, that is the average SDR of mixed speech $y(t)$ with the original target speech $x_1(t)$ and $x_2(t)$ is 0.15. Table~\ref{tab:sdri} lists the average SDRi obtained by TasTas and almost all the results in the past three years, where IRM means the ideal ratio mask
\begin{equation}
M_s=\frac{|X_s(t,f)|}{\sum_{s=1}^{S}|X_s(t,f)|}
\label{eq:irm}
\end{equation}
applied to the STFT $Y(t,f)$ of $y(t)$ to obtain the separated speech, which is evaluated to show the upper bounds of STFT based methods, where $X_s(t,f)$ is the STFT of $x_s(t)$.

\subsection{Results and Discussions}
\label{ssec:results}

In this experiment, TasTas is compared with several classical approaches, such as DPCL~\cite{hershey2016deep}, TasNet~\cite{luo2017tasnet}, Conv-TasNet~\cite{luo2018tasnet}, 
DPRNN-TasNet~\cite{luo2019dual}, Wavesplite~\cite{zeghidour2020wavesplit}, and Nachmani's~\cite{nachmani2020voice}. Use notation TasTas(I, $x_1$, $x_2$, ... , $x_n$) to denote our prosposed system with speaker \textbf{I}dentity-aware dual-path  BiLSTM, and $x_1$  dual-path BiLSTM blocks in the first stage, $x_2$ blocks in the second stage, etc.. If there is no 'I' in TasTas, it uses ordinary SI-SDR loss in training. Thus DPRNN-TasNet is just TasTas(6).

Table~\ref{tab:sdri} lists the results obtained by our methods and almost all the results in the past four years, where IRM means the ideal ratio mask. Compared with these baselines, TasTas obtained an absolute advantage, once again surpassing the performance of stage-of-the-art. TasTas has achieved the most significant performance improvement compared with baseline systems, and it breaks through the upper bound of STFT based methods a lot (more than 7.5dB).  

For  the  \textbf{ablation} study,  Table~\ref{tab:sdri} shows that TasTas(I, 6, 6) is about 0.3dB better than TasTas(6, 6) in SDRi, and TasTas(6, 6) is 0.68dB better than TasTas(6) in SDRi. That means both the  speaker identity-aware dual-path BiLSTM and the iterative multi-phase decontaminated scheme are effective in boost the performance.

\begin{table}[th]
\caption[sdri]{SI-SDRi(dB), SDRi(dB), PESQ, and ESTOI(\%)  in a comparative study of different state-of-the-art separation methods on the WSJ0-2mix dataset.  \textbf{SF}  stands for  TasTas.}\label{tab:sdri}
\centering
\begin{tabular}{c|c|c|c|c}
\hline
\hline
Method & SI-SDRi & SDRi & PESQ & ESTOI\\
\hline
\hline
DPCL~\cite{hershey2016deep}  & - & 5.9  & - & -\\
uPIT-BLSTM~\cite{yu2017permutation}  & - & 10.0 & 2.84 & - \\
ADANet~\cite{luo2018speaker}  & - & 10.5& 2.82 & - \\
DPCL++~\cite{isik2016single} & - & 10.8 & - & - \\
TasNet~\cite{luo2017tasnet} & - & 11.2 & - & -\\
FurcaX~\cite{shi2019furcax} & - & 12.5 & - & - \\
IRM & - &13.0 &  3.68  & 92.9 \\
Wang et al.~\cite{wang2019deep}  & - & 15.4 & 3.45 & - \\
Conv-TasNet~\cite{luo2019conv-tasnet}   & 15.3 & 15.6 & 3.24 & -\\
Deep CASA~\cite{liu2019divide} & 17.7 & 18.0 & 3.51 & 93.2\\
FurcaNeXt~\cite{zhang2020furcanext} & - & 18.4 & - & -\\
DPRNN-TasNet~\cite{luo2019dual} & 18.8 & 19.0 & - & - \\
Wavesplite~\cite{zeghidour2020wavesplit} & 19.0 & 19.2 & - & - \\
Nachmani's~\cite{nachmani2020voice} & 20.12 & - & - & - \\
\hline
\hline
  TasTas(6 ,6)  (ours)   & 19.47 & 19.68 & 3.62 &  94.01 \\
  TasTas(I, 6, 6)  (ours)   & 19.76 &  19.96  & 3.64 & 94.19 \\
  TasTas(8, 9)  (ours)   & 20.35 & 20.55 & 3.69 &  94.86 \\
\hline
\end{tabular}
\end{table}
\section{Conclusion}
\label{sec:conclusion}

In this paper, we investigated the effectiveness of dual-path BiLSTM block based modeling for multi-talker monaural speech separation. We propose TasTas do to speech separation. Benefits from the strength of end-to-end processing, dual-path BiLSTM, speaker identity consistency loss, and the multi-stage elaborated iterative scheme, the best performance of  TasTas achieve the new state-of-the-art of 20.55dB SDRi on the public WSJ0-2mix data corpus.

\section{Acknowledgements}
We would like to thank Dr. Nachmani at Tel-Aviv University \& Facebook AI Research valuable discussions on the training of ID-Net.

\bibliographystyle{IEEEtran}
\bibliography{furcanext_arxiv}

\end{document}